\documentclass[aps,pra,twocolumn]{revtex4-2}
\usepackage{graphicx}

\begin{document}

\title{Spontaneous strain in quasi-two-dimensional \\
Janus CdSe nanoplatelets and its microscopic mechanisms}


\author{Alexander I. Lebedev}
\email{swan@scon155.phys.msu.ru}
\affiliation{Physics Department, Moscow State University, 119991 Moscow, Russia}

\date{\today}

\begin{abstract}
Spontaneous strain and spontaneous folding of thin nanoplatelets are
known phenomena whose microscopic mechanisms are still debating. In this work,
first-principles calculations are used to study the mechanical stresses that
arise in Janus CdSe nanoplatelets and result in their spontaneous strain.
Calculations reveal the existence of three microscopic mechanisms of this
phenomenon. Two bulk mechanisms are associated with the inverse piezoelectric
effect in an electric field created by the difference in electronegativities
of ligands and by the depolarizing field resulting from the difference in the
potential jumps in electrical double layers on the surfaces of nanoplatelets.
These mechanisms account for 5--25\% of the observed effect. The third
mechanism is associated with the surface strain of nanoplatelets by bridging
bonds, and its influence is predominant. It is shown that the latter mechanism cause
spontaneous folding of thin CdSe nanoplatelets and, depending on the values
of surface stresses and lateral orientation of nanoplatelets, can result
in formation of their experimentally observed structures such as scrolls,
spirals, and twisted ribbons.

\texttt{DOI: 10.1021/acs.jpcc.3c01934 (J. Phys. Chem. C 127, 9911 (2023))}
\end{abstract}


\maketitle

\section{Introduction}

Colloidal semiconductor nanoplatelets have been the topic of intense research
in recent years~\cite{ChemMater.25.1262,ChemRev.116.10934,ChemRev.2c00436}. Among
them, nanoplatelets of cadmium chalcogenides have attracted much attention
due to their outstanding optical properties. They are direct band gap semiconductors.
Their perfectly defined thickness and strong quantum confinement in one direction
result in extremely narrow emission and absorption lines~\cite{NatureMater.10.936}.
Furthermore, they display giant oscillator strength~\cite{PhysRevB.91.121302},
high quantum yield~\cite{ChemMater.25.1262}, short radiative recombination
times, and very low lasing thresholds~\cite{NanoLett.14.2772,NatureNanotechnol.9.891},
which can be used in a variety of optoelectronic applications.

An interesting feature of thin semiconductor nanoplatelets is their tendency
to spontaneous folding~\cite{CommunChem.5.7}. This phenomenon has been most
thoroughly studied in II--VI semiconductors. For example, in thin
(2--4~monolayers) CdSe and CdTe nanoplatelets with zinc-blende structure
and lateral sizes larger than 100~nm, it results in formation of
scrolls~\cite{JAmChemSoc.130.16504, JAmChemSoc.134.18591,ChemMater.25.639,
QuantElectron.45.853,ChemMater.29.579,ChemMater.30.1710,ChemMater.31.9652,
NanoLett.19.6466,ACSNano.16.2901}, spirals (helicoids)~\cite{NanoLett.14.6257,
NanoLett.19.6466, ACSNano.13.5326,CommunChem.5.7} as well as twisted
ribbons~\cite{NanoLett.19.6466,SciAdv.3.e1701483}. The diameter of the scrolls
and spirals varies from 20 to 100~nm. Spontaneous folding slightly shifts the
exciton absorption and luminescence bands to longer wavelengths, but can
significantly reduce the luminescence quantum yield. An interesting feature of
the above spiral and twisted structures is that they have chiral optical
properties~\cite{SciAdv.3.e1701483,ACSNano.11.7508,ACSNano.16.2901}. The
occurrence of spontaneous folding has been attributed to an asymmetry of
mechanical stresses created by surface ligands~\cite{JAmChemSoc.134.18591,
ChemMater.25.639,ChemMater.30.1710,ACSNano.16.2901}, but its microscopic
mechanism still remains questionable if one takes into account the perfect
geometric symmetry of the semiconductor part of nanoplatelets.

\begin{figure}
\centering
\includegraphics{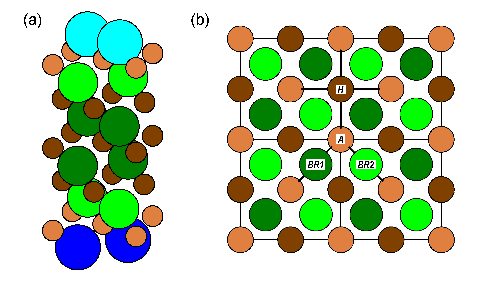}
\caption{\label{fig1}(a) Structure of a 4~ML thick Janus CdSe nanoplatelet
covered with two different monoatomic ligands (shown in cyan and blue).
(b) Possible positions of ligands on the (001) surface of CdSe: atop ($A$),
bridge(1) ($BR1$), bridge(2) ($BR2$), and hollow ($H$). Thick lines show the
chemical bonds. The surface Cd atoms are shown in orange, other Cd atoms are
brown, the Se atoms closest to surface are in green, and other Se atoms are
dark green.}
\end{figure}

The aim of this work is a theoretical study of the phenomenon of spontaneous
strain that arises in CdSe nanoplatelets when their two surfaces are covered
with different ligands (Fig.~\ref{fig1}a). Such nanoplatelets are called the
Janus ones. In contrast to true Janus nanoplatelets like MoSSe~\cite{NatureNanotechnol.12.744}
in which the real
atomic substitution takes place, our Janus CdSe nanoplatelets retain their atomic
structure, and the only change that is produced is the substitution of ligands.
To the best of our knowledge, this is the first study of CdSe Janus
nanoplatelets. Our interest to these objects stems from the fact that the
strain that arises in them may be one of the cause of spontaneous folding of
nanoplatelets. At the end of this article, experimental data supporting this
hypothesis will be presented.

In this work, the microscopic mechanisms that result in spontaneous strain and
spontaneous folding of Janus CdSe nanoplatelets will be established. The
obtained results clarify the nature of physicochemical phenomena occurring
at the interface between the nanoplatelet and surface ligands and show
the ways to control the above phenomena.

\section{Computational Details}

The objects of this study are the (001)-oriented CdSe nanoplatelets with
zinc-blende structure. As was established in previous studies,~\cite{ChemRev.2c00436,ChemMater.31.9652,JPhysChemC.125.6758}
these nanoplatelets contain one excess atomic layer of cadmium, and
two extra electrons from each Cd atom enter the conduction band. In order to
bound these electrons and retain the semiconductor properties, two $X$-type
ligands should be placed on both sides of the nanoplatelet.

The first-principles DFT calculations of the equilibrium geometry of nanoplatelets
were carried out using the ABINIT software~\cite{abinit3}. The local density
approximation (LDA) and norm-conserving optimized pseudopotentials~\cite{PhysRevB.41.1227}
constructed using the OPIUM program~\cite{opium} were used in the calculations.
The plane wave cutoff energy was 30~Ha (816~eV), the integration over the Brillouin
zone was carried out on a 8$\times$8$\times$8 Monkhorst--Pack mesh for bulk
CdSe and on a 8$\times$8$\times$1 Monkhorst--Pack mesh for nanoplatelets.
The lattice parameters and the positions of atoms were relaxed until the forces
acting on the atoms became less than $5 \times 10^{-6}$~Ha/Bohr (0.25~meV/{\AA})
while an accuracy of the total energy calculation was better than 10$^{-10}$~Ha.

The calculated lattice parameter of bulk zinc-blende CdSe was $a = 6.0175$~{\AA},
in good agreement with the experimental data of 6.077~{\AA}~\cite{Adachi2009}
if one takes into account a small underestimation of the lattice parameter
typical of the LDA.
Under the tetragonal distortion, the FCC Bravais lattice of bulk CdSe transforms
to body-centered tetragonal lattice, which loses some symmetry elements and
transforms to a primitive tetragonal lattice when forming nanoplatelets.
In this structure, the translation vectors are rotated by 45$^\circ$ relative to
the axes of the original cubic cell, and the lattice parameters in the $xy$ plane
are $a_0 = b_0 \approx a/\sqrt{2}$. Our computations were performed on supercells,
into which a vacuum gap of $\sim$20~{\AA} in the $z$~direction normal to the
$xy$~plane was added to isolate the nanoplatelets from each other.%
    \footnote{In reality, the nanoplatelet under consideration has a symmetry of the
    $p{\bar 4}m2$ layer group. The possibility of its modeling using 3D modeling
    programs within the $P{\bar 4}m2$ space group was discussed in Ref.~\citenum{JPhysChemC.125.6758}.}
When performing the structure optimization ($a_0$, $b_0$, and the positions of atoms),
the $c_0$ supercell lattice parameter remained fixed.

\section{Results and discussion}

Two groups of causes that may result in appearance of spontaneous strain in
Janus structures are expected. First, these are two bulk effects. The first of
them is due to the difference in electronegativities of ligands, which results
in a redistribution of the electron density and the appearance of (1) a vertical
electric field in the nanoplatelet and (2) a shear strain of the supercell as
a result of the inverse piezoelectric effect (we recall that bulk CdSe is a
piezoelectric~\cite{Adachi2009}). The second bulk effect is associated with
the appearance of a vertical depolarizing field resulting from the difference
in the potential jumps in the electrical double layers which appear when
the positively charged nanoplatelet is covered with two different negatively charged
$X$-type ligands at two sides. The second group of causes are various surface
effects arising from mechanical stresses that appear upon formation of chemical
bonds with ligands.

Because the effect of real ligands on the structure of nanoplatelets includes
many aspects and may be very complex, the simplest monoatomic and diatomic ligands
were used in our \emph{ab initio} calculations to identify the mechanisms
by which they influence the structure. This gives us a basis to classify the
effects produced by more complex ligands.

\subsection{Monoatomic ligands}

\begin{table}
\caption{\label{table1}Relative per-atom adsorption energies for F, Cl, and Br
ligands in different positions on the (001) surface of a 4~ML thick CdSe nanoplatelet
and the corresponding equilibrium lattice parameter. The energy of the structure
with the lowest energy is taken as the reference point.}
\begin{ruledtabular}
\begin{tabular}{ccccccc}
              & \multicolumn{2}{c}{F atoms} & \multicolumn{2}{c}{Cl atoms} & \multicolumn{2}{c}{Br atoms} \\
Configuration & $E$~(eV) & $a_0$~({\AA})    & $E$~(eV) & $a_0$~({\AA})     & $E$~(eV) & $a_0$~({\AA}) \\
\hline
bridge(1) & 0.0   & 4.1457 & 0.0   & 4.1993 & 0.0   & 4.2157 \\
bridge(2) & 0.864 & 3.9289 & 0.766 & 4.0142 & 0.724 & 4.0521 \\
atop      & 1.099 & 4.2425 & 0.959 & 4.2104 & 0.909 & 4.2105 \\
hollow    & 1.311 & 3.8322 & 0.818 & 3.9205 & 0.699 & 3.9638 \\
\end{tabular}
\end{ruledtabular}
\end{table}

We start our study of the surface monoatomic ligands with the determination of their
energetically most favorable positions. Different possible positions of adsorbed
atoms on the (001) surface of zinc-blende structure are shown in Fig.~\ref{fig1}b.
It should be noted that in this structure there are two different types of
bridging positions, bridge(1) and bridge(2). This follows from the fact that
covalent chemical bond in zinc-blende structure is based on $sp^3$ hybridized
orbitals, and as a result of its directionality, the adsorption energies of atoms
in these two positions are different. The relative adsorption energies of F, Cl,
and Br atoms on the surface of a CdSe nanoplatelet are given in Table~\ref{table1}.
As was noted earlier~\cite{PhysRevB.95.165414}, the most energetically favorable
position of the ligand is the position, in which its local structure reproduces
the structure of bulk CdSe. All other positions, including the bridge(2) one,
have significantly higher energies. As follows from the obtained results, all
considered ligands induce the in-plane contraction of the structure (in bulk
CdSe, the calculated $a_0 = a/\sqrt{2}$ value is 4.2550~{\AA}).

To separate the surface effects from the bulk ones, the dependence of stresses
created in nanoplatelets on their thickness $N$ (given in monolayers) was studied.
We focused our attention on stresses in order to exclude from consideration the
elastic properties of nanoplatelets because they change with thickness. For this
purpose, the calculations were carried out for nanoplatelets clamped on a square
substrate ($a_0 = b_0$) with the lattice parameter satisfying the boundary conditions
$\sigma_1 + \sigma_2 = 0$ (hereinafter, indices are given in Voigt notation).
The parameter of interest to us was the stress $\sigma_1 = -\sigma_2 \ne 0$. The
obtained results are given in Table~\ref{table2}.%
    \footnote{The boundary conditions depend on the orientation of the unit
    cell. For example, for the unit cell with the [$a/2$,$a/2$,0], [$-a/2$,$a/2$,0],
    [0,0,$c_0$] translation vectors, the boundary condition are
    $\sigma_1 = \sigma_2 = 0$, and the stress in question is $\sigma_6$.}

\begin{table}
\caption{\label{table2}Stress $\sigma_1$ in Janus CdSe nanoplatelets of
various thicknesses~$N$ covered with F--Cl and Cl--Br ligand pairs. To unambiguously
determine the sign of the stress, the more electronegative ligand atom in the
nanoplatelet was always chosen
to be the first one. The last column presents the $a_0$ and $b_0$ lattice
parameters of the relaxed structure whose translation vector~${\bf R}_1$ is directed
along the bridging bond for a larger ligand.}
\begin{ruledtabular}
\begin{tabular}{cccccc}
$N$ & $c_0$  & $\sigma_1$ (GPa) & $\sigma_1 c_0$ & $a_0 = b_0$ ({\AA}) & $a_0$,~$b_0$ ({\AA}) \\
    & (Bohr) &                  &                & (clamped)           & (relaxed) \\
\hline
\multicolumn{6}{c}{F and Cl ligands} \\
2   & 66.0  & $-0.3235 $ & 21.35 & 4.1139 & 4.2138, 4.0333 \\
3   & 66.0  & $-0.3276 $ & 21.62 & 4.1509 & 4.2279, 4.0856 \\
4   & 66.0  & $-0.3295 $ & 21.75 & 4.1729 & 4.2350, 4.1179 \\
5   & 72.0  & $-0.3032 $ & 21.83 & 4.1867 & 4.2392, 4.1396 \\
6   & 78.0  & $-0.2805 $ & 21.88 & 4.1968 & 4.2420, 4.1553 \\
\multicolumn{6}{c}{Cl and Br ligands} \\
2   & 66.0  & $-0.0794 $ & 5.24  & 4.1759 & 4.2003, 4.1518 \\
3   & 66.0  & $-0.0792 $ & 5.23  & 4.1954 & 4.2138, 4.1769 \\
4   & 66.0  & $-0.0793 $ & 5.23  & 4.2072 & 4.2224, 4.1925 \\
5   & 72.0  & $-0.0727 $ & 5.24  & 4.2155 & 4.2281, 4.2031 \\
6   & 78.0  & $-0.0672 $ & 5.24  & 4.2212 & 4.2320, 4.2105 \\
\end{tabular}
\end{ruledtabular}
\end{table}

When analyzing the obtained data, it should be taken into account that in
the ABINIT program the stress is calculated as a force acting on the total
surface area, which in structures containing a vacuum gap includes the
area occupied by this layer \cite{JApplPhys.124.164302}. This is why, when
comparing data obtained on supercells of different lengths, it is more correct
to use the product of the stress and the full period of the supercell in the
$z$~direction, $\sigma_1 c_0$. As follows from Table~\ref{table2}, this value
weakly depends on $N$ and demonstrates a gradual saturation as the thickness of
the nanoplatelet increases. A monotonous increase in the lattice parameter in
clamped structures with increasing~$N$ confirms our earlier conclusion that
all considered monoatomic ligands contract the structure.

Similar results were obtained for clamped CdSe Janus structures with surface
Cl and Br ligands (Table \ref{table2}).

When the $a_0 = b_0$ constraint is released, the structure
relaxes and a noticeable spontaneous strain in the [110] direction of the original
cubic structure appears.%
    \footnote{In our supercell setting, the relaxation of the structure produced
    by the stress $\sigma_1 = -\sigma_2 \ne 0$ results in a change of lengths
    of the translation vectors, $a_0 \ne b_0$.}
According to experiment, this direction coincides with the direction of spontaneous
folding of nanoplatelets~\cite{ChemMater.25.639,CommunChem.5.7} and their
spontaneous strain~\cite{ChemMater.31.9652}. The in-plane lattice parameters $a_0$
and $b_0$ of the orthorhombic relaxed structure are given in Table~\ref{table2}.
It is seen that as the thickness of the nanoplatelets increases, the magnitude
of spontaneous strain decreases, which is explained by an increase in
the mechanical rigidity of nanoplatelets (we note, however, that the elastic
modulus $C_{1212}$ is not a linear function of $N$).

Judging from a very weak dependence of $\sigma_1 c_0$ on the thickness of
nanoplatelets, the perturbation source is localized. Unfortunately, the problem
of separating the surface and bulk effects is complicated by the fact that in
our geometry the Poisson's equation for the electric field is one-dimensional,
and in the absence of screening (we are dealing with a semiconductor), the
vertical electric field in the structure remains uniform. This is a direct
consequence of the divergence theorem. In a long supercell, when the thickness
of a semiconductor part of nanoplatelet changes, the lateral \emph{pressure}
produced by bulk electric field decreases, but the lateral \emph{surface area}
occupied by the semiconductor part increases, thus compensating each other.
This makes it impossible to decide whether the observed spontaneous strain
is caused by surface or bulk effects.

To separate these effects, additional calculations for two hypothetical structures
were performed. The first of these was the nanoplatelet clamped on a square
substrate with the atop position of both terminating atoms. In this configuration,
the bulk effect associated with the difference in electronegativities of ligands
should remain practically unchanged (ultimately, it is determined by the
difference in energies of partially filled $p$ orbitals in the ligands), and an
anisotropic surface stresses that occur when the ligands are in bridge positions
should disappear due to symmetry. For nanoplatelets with a thickness of 4~ML and
surface atop F and Cl ligands, the obtained $\sigma_1$ value was $-$0.0189~GPa,
which was about 20 times less than in the configuration with bridging bonds
(Table~\ref{table2}). For nanoplatelets with a thickness of 3~ML and 5~ML, the
calculations gave the values $\sigma_1 = -0.0175$~GPa and $\sigma_1 = - 0.0172$~GPa,
respectively. The sign of the effect remained the same as in the bridge(1)
configuration. For a 4~ML thick nanoplatelet with atop Cl and Br ligands, the
result was similar: the sign of the effect remained the same as in the bridge(1)
configuration, but its magnitude was an order of magnitude smaller
($-$0.0060~GPa).

The second hypothetical structure was a clamped nanoplatelet with hollow positions
of both ligands. In this structure, at a fixed thickness of the nanoplatelet,
the vertical electric field resulting from the difference in electronegativities
should remain almost the same as in the bridge(1) configuration, but the anisotropy
of surface stresses should disappear due to symmetry. Calculations for a
4~ML thick nanoplatelets with the hollow positions of F and Cl ligands gave
the value $\sigma_1 = -0.0620$~GPa, which turned out to be 3~times stronger than
for the atop configuration. For nanoplatelets with a thickness of 3~ML and 5~ML,
the $\sigma_1$ values were $-$0.0355~GPa and $-$0.0746~GPa, respectively. The
fact that the $\sigma_1$ values in structures with atop and hollow positions are
noticeably different indicates the existence of several competing bulk mechanisms.
In our opinion, the second bulk mechanism is associated with a depolarizing
electric field caused by different potential jumps in electrical double layers
on two surfaces of nanoplatelets.

In an electrically neutral nanoplatelet, the positive charge of the semiconductor
part is compensated by the negative charge of $X$ ligands. In symmetric
nanoplatelets, the potential jumps created by electrical double layers on two
surfaces compensate each other and the bulk electric field is zero. However, in
Janus nanoplatelets, these dipole moments do not compensate each other anymore,
and an additional vertical electric field arises in the structure. This field,
due to the inverse piezoelectric effect, creates additional mechanical stresses
in the nanoplatelet. For example, in the 4~ML thick CdSe nanoplatelet, the
distance between the layer of bridging~F and Cl atoms and the nearest layers of
cadmium atoms are 0.689 and 1.352~{\AA} for the bridge(1) positions, 1.934 and
2.269~{\AA} for atop positions, and 0.619 and 1.075~{\AA} for the hollow positions.
The corresponding differences in the interplanar Cd--F and Cd--Cl distances are
0.663, 0.335, and 0.456~{\AA}. A qualitative analysis of the distribution of fields
created by the difference in electronegativities and by dipole moments on the surfaces
of the nanoplatelet shows that the contribution of dipole moments should increase
in the series atop--hollow--bridge(1), and its sign should coincide with the
sign of the effect produced by the difference in electronegativities. This is
exactly what is observed in our calculations.

The difference in the potential jumps on two sides of Janus nanoplatelets
is clearly seen in the averaged electrostatic potential curves calculated for a
special supercell containing two vacuum gaps and two oppositely oriented Janus
nanoplatelets to suppress the depolarizing field (see Fig.~\ref{figS1} in the
Supporting information). The potential curve
was obtained using the macroaveraging technique~\cite{PhysRevLett.61.734}
implemented in the \texttt{macroave} program of the ABINIT software package.
It is seen that the difference in the potential jumps can be as large as 0.77~eV.

\begin{figure}
\centering
\includegraphics{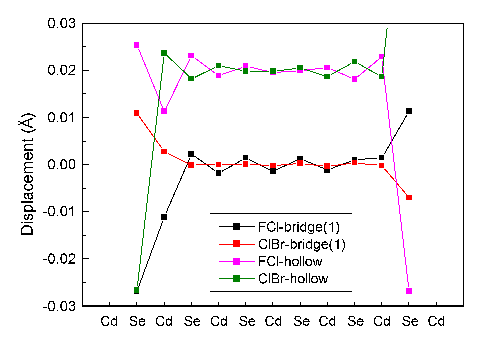}
\caption{\label{fig2}Deviation of the $z$~coordinates of individual atoms from
the half-sum of the $z$~coordinates of neighboring atoms in clamped 6~ML thick
Janus CdSe nanoplatelets with F--Cl and Cl--Br ligand pairs (the more
electronegative atom is located on the left). Two upper curves corresponding
to the same ligand pairs in the hollow position are shifted by 0.02~{\AA} for
clarity.}
\end{figure}

Displacements of atoms in a nanostructure from their ideal positions may provide
additional information on the discussed phenomena. If one suppresses the
strain by fixing the structure on a square substrate, the appearance of such
displacements can indicate the presence of electric and stress fields in the
structure. For example, the existence of a vertical electric field in clamped
nanoplatelets with the bridge(1) position of ligands is confirmed by the
observation of spatially uniform oscillations of interplanar distances
(Fig.~\ref{fig2}). Since the chemical bond in CdSe is partially ionic, this
field acts on small positive and negative charges of atoms and results in
out-of-phase displacements of the planes occupied by neighboring atoms.
The direction of atomic displacements exactly corresponds to the direction
of the electric field created by the difference in electronegativities.

When ligands are moved to the atop and hollow positions, the character of
the displacements changes. In particular, in a number of structures one can
observe damped oscillations whose phase changes by 180$^\circ$ when passing
to the other side of the nanoplatelet. To better understand the nature of
the observed displacements, additional calculations were carried out for
both conventional and Janus nanoplatelets made of silicon, a material in which
the piezoelectric effect is absent. As follows from Fig.~\ref{figS2} of the
Supporting information, similar displacement patterns were observed in
fully symmetric Ge-terminated silicon nanoplatelets. Studies of Janus Si nanoplatelets
with Ge and C terminating atoms show that these oscillations are induced
by the nearest surface, and the sign of the displacements is determined
by the sign of strain (compression or tension) at this surface.
Therefore, oscillations of atomic
displacements observed in CdSe nanoplatelets with a phase change by
180$^\circ$ point to the surface relaxation as the cause of this phenomenon.
However, electrical effects cannot be completely ruled out since the surface
piezoelectricity---the appearance of fields of piezoelectric origin in
differently relaxed layers on the surface---can also play a certain role in
CdSe nanoplatelets.

When the $a_0 = b_0$ constraint is released, the structures with atop and
hollow positions of ligands relax in the same way as the structures with the
bridge(1) position, in accordance with the sign of $\sigma_1$ in clamped
structures. An analysis of the equilibrium geometry of the structures under
discussion shows that the position of monoatomic ligands on bridging bonds
is always symmetric, and the Cd--$X$ bond lengths systematically increase in
the series $X$ = F, Cl, Br. This means that stress-free nanoplatelets
always elongates in the direction of the bridging bond with a larger ligand.

The question can arise if the observed effects are entirely associated with
the bulk electric field produced by the difference in the potential jumps at two
surfaces of the nanoplatelet. Unfortunately, the dipole correction technique~\cite{PhysRevB.59.12301}
is not implemented in the ABINIT package. In order to answer this question, we
performed additional calculations for a number of doubled supercells containing
two vacuum gaps and two oppositely oriented Janus nanoplatelets. It appeared
that in these dipole-corrected calculations, the values of $\sigma_1$ stress
in clamped nanoplatelets are changed by no more than 2\%. At the same time,
the atomic displacements of Fig.~\ref{fig2} become weaker (see Fig.~\ref{figS3}
in the Supporting information) and remind the atomic displacements patterns
produced by the surface relaxation. The calculated profiles of the electrostatic
potential show that without the dipole correction, the depolarizing electric
field can reach 3~MV/cm in the vacuum layer and 200~kV/cm in the semiconductor
part. In the supercells with suppressed depolarizing field, the weaker effect
of the electric field originating from the difference in electronegativities of
ligands is observed (see Fig.~\ref{figS4} in the Supporting information): its
strength is about 30~kV/cm. A parabolic change of the potential in the middle
of the nanoplatelet is due to homogeneous distribution of positive charge in the
semiconductor part. It supports our model of the nanoplatelet as a positively
charged semiconductor nanoplatelet covered with negatively charged $X$ ligands.

To sum up this section, the performed calculations unambiguously indicate
the existence
of three microscopic mechanisms of spontaneous strain in Janus nanoplatelets. Two
bulk mechanisms are associated with the inverse piezoelectric effect which occurs
in an electric field created by different electronegativities of ligands and by
different dipole moments of electrical double layers formed on two surfaces. These
mechanisms account for 5--25\% of the observed effect. The third mechanism is
associated with the surface strain of nanoplatelets by bridging bonds, and its
effect is predominant.

\subsection{Diatomic ligands}

To better understand the mechanism of the surface strain, we now consider
diatomic OH, SH, SeH, and TeH ligands in the bridge(1) position. It was shown
that this position is the equilibrium one since structures with general starting
positions of atoms relax into it. We start with the influence of geometric size
of ligands on the magnitude of stresses in CdSe nanoplatelets clamped on
a square substrate ($a_0 = b_0, \sigma_1 + \sigma_2 = 0$). The results of
calculations for a 4~ML thick nanoplatelet and various ligands paired with more
electronegative F and Cl ligands are given in Table~\ref{table3}.

\begin{table}
\caption{\label{table3}Stress $\sigma_1$ in 4~ML Janus CdSe nanoplatelets
covered with various ligands. The translation vector~${\bf R}_1$ is directed
along the bridging bond of the $Y$~ligand.}
\begin{ruledtabular}
\begin{tabular}{ccccc}
$Y$    & $\sigma_1$ (GPa) & $R_{{\rm Cd}-Y}$ ({\AA}) & $a_0 = b_0$ ({\AA}) & $a_0$,~$b_0$ ({\AA}) \\
       &                  &                          & (clamped)           & (relaxed) \\
\hline
\multicolumn{5}{c}{F and $Y$ ligands} \\
OH     & $+0.0596 $ & 2.219 & 4.1126 & 4.1029, 4.1228 \\
SH     & $-0.4021 $ & 2.504 & 4.1625 & 4.2333, 4.0931 \\
SeH    & $-0.5162 $ & 2.618 & 4.1800 & 4.2733, 4.0924 \\
TeH    & $-0.6336 $ & 2.749 & 4.2130 & 4.3278, 4.1114 \\
\hline
Cl     & $-0.3295 $ & 2.486 & 4.1729 & 4.2350, 4.1179 \\
Br     & $-0.4108 $ & 2.601 & 4.1817 & 4.2602, 4.1141 \\
\hline
COO    & $-0.5846 $ & 2.190\textsuperscript{\emph{a}} & 4.2039 & 4.3089, 4.1110 \\
\hline
\multicolumn{5}{c}{Cl and $Y$ ligands} \\
OH     & $+0.3918 $ & 2.224 & 4.1387 & 4.0755, 4.2141 \\
SH     & $-0.0644 $ & 2.506 & 4.1882 & 4.1997, 4.1757 \\
SeH    & $-0.1767 $ & 2.620 & 4.2055 & 4.2378, 4.1717 \\
TeH    & $-0.2909 $ & 2.750 & 4.2378 & 4.2904, 4.1844 \\
\hline
F      & $+0.3295 $ & 2.197 & 4.1729 & 4.1179, 4.2350 \\
Br     & $-0.0793 $ & 2.603 & 4.2072 & 4.2224, 4.1925 \\
\end{tabular}
\end{ruledtabular}

\textsuperscript{\emph{a}}In contrast to the results of Ref.~\citenum{JAmChemSoc.141.15675},
our calculations show that the most stable configuration of the HCOO ligand is
its tilted configuration. The table presents the shortest Cd--O distance.
\end{table}

It is seen that when passing from OH to larger ligands, the $\sigma_1$ stress
changes its sign from positive to negative, and its negative values continue to
increase with increasing ligand size. This fully supports the idea of the surface
stress as a mechanism of the influence of ligands on spontaneous strain. The OH
ligand causes contraction of the structure in the direction of the bridging bond,
whereas all other ligands elongate it in accordance with the bond length
$R_{{\rm Cd}-Y'}$ ($Y' = {}$O, S, Se, Te). Calculations show that the $Y'$~atoms
that occupy the bridge position remain at an equal distance from the Cd atoms,
but the inclination of the $Y'$--H bond, which is accompanied by slight opposite
lateral shifts of the $Y'$ and H~atoms in the direction perpendicular to the
bridging bond, results in a distortion of the structure. It lowers its symmetry
to $Pm$.

\begin{figure}
\centering
\includegraphics{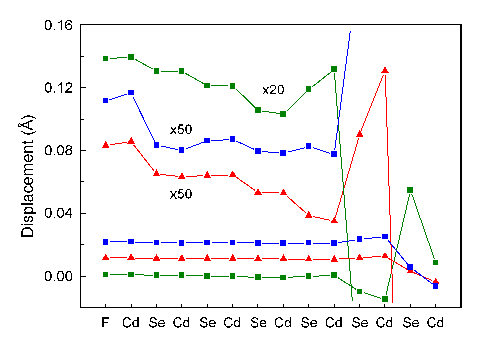}
\caption{\label{fig3}Damping of atomic $y$ displacements in the bulk of a clamped
6~ML thick Janus CdSe nanoplatelet caused by the surface displacement of the
hydrogen atom in OH (green squares), SH (red triangles), and SeH (blue squares)
groups. The top three curves show enlarged displacement patterns inside the
nanoplatelet.}
\end{figure}

The atomic displacements in the $y$~direction produced by distortion of the
surface by different diatomic ligands are shown in Fig.~\ref{fig3}. It is seen
that in the bulk of a semiconductor the atomic displacements rapidly decay
with distance. This indicates the surface origin of the perturbation.
However, the origin of the displacement oscillations deep in structure turns
out to be quite unexpected: for all ligands, an analysis finds a doubled
character of displacements of neighboring atoms, thus indicating the involvement
of TA phonon at the $X$~point of the Brillouin zone, which has a similar
eigenvector, into the relaxation. The observed doubling is typical of
zinc-blende structure, in which vibrations of Cd--Se pairs in the [110] direction
alternately become stretching, then bending, then stretching again.
Weak additional out-of-phase displacements of Cd and Se atoms may indicate the
existence of an electric field parallel to the surface.

Similar results were also obtained for 4~ML thick Janus CdSe nanoplatelets
with OH, SH, SeH, and TeH ligands in the bridge(1) position in a pair with a
more electronegative Cl atom (Table~\ref{table3}).

By comparing data for diatomic ligands with the results obtained for monoatomic
ligands, one can conclude that the surface contribution to spontaneous strain
is determined by the difference in geometric size of ligands. Based on the
lattice parameters of clamped structures at $\sigma_1 + \sigma_2 = 0$
(Table~\ref{table3}),
the ligands can be arranged in the following series according to their strain
effect: OH--F--Cl--SH--Br--SeH--TeH. For all ligands we have considered, the
absolute sign of the strain is negative, which follows from the comparison
of obtained $a_0$ parameters with that calculated for bulk CdSe
($a_0 = a/\sqrt{2} = 4.2550$~{\AA}). Alternatively, the $\sigma_1$ values can be used to
evaluate the effect of ligands on the spontaneous strain.

The geometry of CdSe nanoplatelets is such that regardless of whether they
have an even or odd number of monolayers, the directions of bridging bonds
on their two opposite sides are always perpendicular to each other. When the
monoatomic ligands are the same and we use a 3D modeling program, this guarantees
the preservation of the tetragonal structure. However, in Janus structures, the
symmetry becomes lower. In the case of diatomic ligands, the situation is
additionally complicated by the appearance of a transverse component of atomic
displacements.

Unfortunately, the 3D modeling programs do not allow to change the unit cell
parameters when moving along the~$z$ axis and thus do not allow one to clearly
see what will happen with a free-standing 2D nanoplatelet. However, this
is not hard to imagine. If we change to a curvilinear coordinate system, in
which the $x$ and $y$ axes run parallel to the atomic chains on the surface
and the $z$~axis is directed perpendicular to them, then the mechanism of
surface strain established above will result in a bending of the structure,
and the interatomic distances in chains along the $x$ and $y$~axes will become
$z$-dependent.

Anisotropic compressive stresses directed perpendicular to each other at two
sides of a nanoplatelet will result in a saddle-shaped distortion of Janus
nanoplatelets with small lateral sizes and will be characterized by two radii of
principal curvatures. To be able to roll up nanoplatelets with a large lateral
size into scrolls, one of the curvatures should be close to zero (the experiment
shows that in most CdSe-based nanostructures the product of the two principal
curvatures---so-called Gaussian curvature---is zero~\cite{CommunChem.5.7}). If
so, then the belt- or ribbon-type nanostructures oriented in the [110] direction
will form scrolls, whereas nanostructures oriented in the [100] direction will
form spirals. When the curvatures of [100]-oriented ribbons are close, the
formation of a twisted-ribbon structure becomes possible.

In this work, we did not plan to discuss the properties of nanoplatelets
covered with long-chain organic molecules, since we were mainly interested in
studying the structure of the interface between a semiconductor and ligands
and the mechanisms that govern it. For organic ligands, the problem
becomes more difficult since one of the oxygen atoms of the carboxyl group
that forms a chemical bond occupies a distorted bridge(1) position with three
different Cd--O distances. So, the surface distortions become even more complex
and an additional problem arises, which requires to take into account the strain
interaction in the layer of organic molecules. According to our calculations,
the geometric size of the carboxyl group, judging from its effect on the lattice
parameter, is slightly smaller than that of the TeH ligand (Table~\ref{table3}).

Returning back to the phenomenon of spontaneous folding of nanoplatelets, we
note that one of the puzzles in explaining this phenomenon is why it appears
when both sides of the nanoplatelet are covered with ligands symmetrically.
But if we assume that Janus nanoplatelets spontaneously occur in a
colloidal solution, then an increase in the equilibrium lattice parameter
in these nanoplatelets that arises when a larger ligand is placed on one of its
sides should obviously result in such a folding.%
    \footnote{Bulk effects can only result in spontaneous strain of the
    nanoplatelet without giving rise to its spontaneous folding.}
This may occur, for example, as a result of the ordering of oleic and acetate
groups on two sides of nanoplatelets (according to NMR~\cite{JAmChemSoc.141.15675},
nanoplatelets synthesized using lead acetate usually contain both types of groups).
The stability of such configurations is ensured by the Gorsky effect (diffusional
exchange of ligands in a stress field). Indirect evidences for formation of
Janus structures can be observed in refs.~\citenum{ChemMater.29.579,ChemMater.30.1710,
ChemMater.31.9652,ACSCentSci.5.1017}, in which the distance between layers in
nanoplatelets rolled into scrolls as well as in nanotubes
are markedly less than twice the length of the long-chain organic
stabilizer molecules (about 22~{\AA} for oleic acid and hexadecanethiol).
This may indicate the appearance of of short-chain organic ligands
in combination with the
long-chain ones in the gap. It should also be noted that spontaneous strain
along the [110] direction is typical of Janus nanoplatelets: according to our
calculations, nanoplatelets covered on both sides with the same organic ligands
spontaneously strain in the [100] direction.

To estimate the curvature radius of a rolled nanostructure, one can use the
$a_0$ and $b_0$ values from Table~\ref{table2}, although for a real free-standing
nanoplatelet, the difference between these parameters can be even stronger.
For a 3~ML thick nanoplatelet, the characteristic radius of curvature is
$\sim$36~nm and agrees with experiment.

\section{Conclusions}

The first-principles calculations of the geometry of Janus CdSe nanoplatelets
performed in this work indicate the existence of three microscopic mechanisms
of their spontaneous strain. Two bulk mechanisms are associated with the
inverse piezoelectric effect in an electric field created by the difference
in electronegativities of ligands and in the depolarizing field resulting from
the difference in the potential jumps in electrical double layers on two
surfaces of nanoplatelets. These mechanisms account for 5--25\% of the observed
effect. The third mechanism is due to the surface strain of the nanoplatelet
by bridging bonds, and its effect is predominant. The latter mechanism
initiates spontaneous folding of thin nanoplatelets and, depending on the
magnitude of surface stresses and orientation of nanoplatelets in the lateral
plane, can result in formation of scrolls, spirals, or twisted ribbons.

\appendix
\section{Supporting information}

\setcounter{figure}{0}
\renewcommand{\thefigure}{S\arabic{figure}}

\begin{figure}
\includegraphics{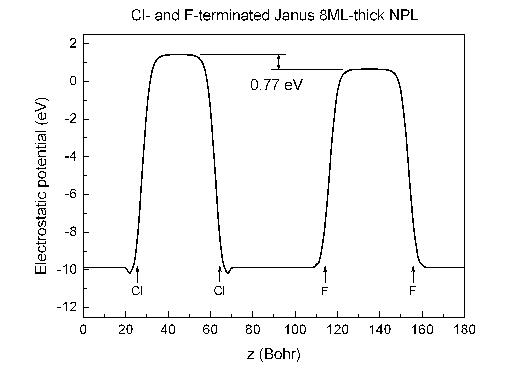}
\caption{\label{figS1}Averaged electrostatic potential in dipole-corrected structure
consisting of two vacuum gaps and two oppositely oriented clamped F- and Cl-terminated
Janus 8~ML-thick CdSe nanoplatelets.}
\end{figure}

Fig.~\ref{figS1} shows the averaged electrostatic potential calculated using the \texttt{macroave}
program from the \texttt{ABINIT} software package for a dipole-corrected structure consisting
of two vacuum gaps and two oppositely oriented clamped F- and Cl-terminated Janus 8~ML-thick
CdSe nanoplatelets. It is seen that in two vacuum layers, the electrostatic potentials
which result from the difference in the potential jumps in electrical double layers at the
surfaces of each nanoplatelet, differ by 0.77~eV. In structures without the dipole correction,
this difference in the potential jumps was shown to induce the depolarizing fields up to
3~MV/cm in the vacuum region and up to 200~kV/cm in the semiconductor part.

\begin{figure}
\includegraphics{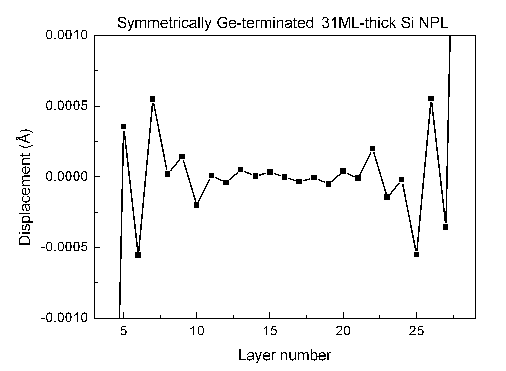}
\caption{\label{figS2}Deviations of the $z$~coordinates of individual atoms from
the half-sum of the $z$~coordinates of neighboring atoms in clamped 31~ML-thick
silicon nanoplatelet terminated with Ge atoms.}
\end{figure}

The deviations of the $z$~coordinates of individual atoms from the half-sum of the
$z$~coordinates of neighboring atoms in a clamped 31~ML-thick silicon nanoplatelet
symmetrically terminated with Ge atoms are shown in Fig.~\ref{figS2}. One can see an interference
of damped oscillations which have a phase shift of 180$^\circ$ and come from two opposite
sides of the nanoplatelet. The origin of this structure is the surface relaxation.

\begin{figure}
\includegraphics{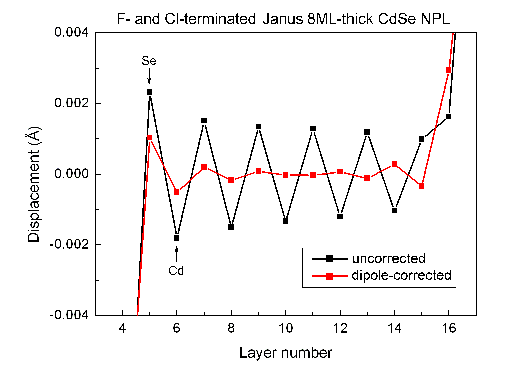}
\caption{\label{figS3}Deviations of the $z$~coordinates of individual atoms from
the half-sum of the $z$~coordinates of neighboring atoms in clamped 8~ML-thick
Janus CdSe nanoplatelet terminated with F and Cl ligands before and after the
dipole correction.}
\end{figure}

Fig.~\ref{figS3} shows the effect of the dipole correction on the deviation of the $z$~coordinates
of individual atoms from the half-sum of the $z$~coordinates of neighboring atoms in a
clamped 8~ML-thick Janus CdSe nanoplatelet terminated by F and Cl ligands. It is seen that
without the dipole correction, the depolarizing field induces, via the inverse piezoelectric
effect, large out-of-phase shifts of Cd and Se atoms. When the depolarizing field
are switched off by using a special supercell with two vacuum gaps and two oppositely
oriented Janus nanoplatelets, the oscillations become much weaker and display a picture
of two interfering dumped waves coming from two opposite sides of the nanoplatelet and
having the 180$^\circ$ phase shift typical of the surface relaxation (Fig.~\ref{figS2}).

\begin{figure}
\includegraphics{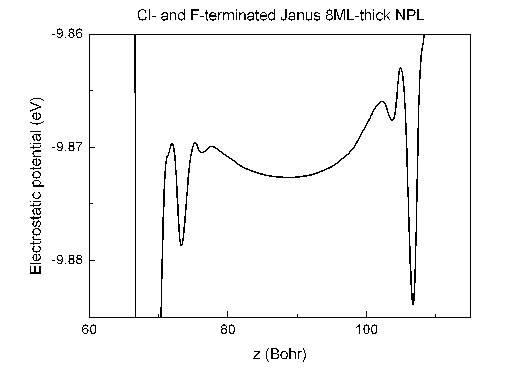}
\caption{\label{figS4}Enlarged portion of Fig.~\ref{figS1} showing the profile of the averaged
electrostatic potential inside the nanoplatelet.}
\end{figure}

Fig.~\ref{figS4} presents an enlarged portion of the averaged electrostatic potential of Fig.~\ref{figS1}
inside the nanoplatelet. As the nanoplatelet is mechanically clamped and the depolarizing
field is fully compensated, the only remaining perturbation is the electric field induced
by the difference in electronegativities of ligands. It produces a change of the averaged
electrostatic potential with a slope of $\sim$30~kV/cm. A parabolic change of the potential
in the middle of the nanoplatelet is due to homogeneous distribution of positive charge
in the semiconductor part. It supports its model as a positively charged semiconductor
nanoplatelet covered with negatively charged $X$ ligands.

\clearpage

\begin{acknowledgments}
The work was financially supported by Russian Science Foundation, grant
22-13-00101. The author thanks R.B. Vasiliev who stimulated the author's
interest to nanoplatelets for numerous fruitful discussions of various
aspects of the problem.
\end{acknowledgments}

\providecommand{\BIBYu}{Yu}

\end{document}